\documentclass[aps,prl,twocolumn,groupedaddress]{revtex4}

\usepackage{graphicx}
\usepackage{dcolumn}
\usepackage{bm}

\begin{document}

\title{\large  Proximity effects at the interface of a superconductor and a topological insulator\\
in NbN-$\rm Bi_2Se_3$ thin film bilayers.}

\author{G. Koren}
\email{gkoren@physics.technion.ac.il} \affiliation{Physics
Department, Technion - Israel Institute of Technology Haifa,
32000, ISRAEL} \homepage{http://physics.technion.ac.il/~gkoren}

\date{\today}
\def\bfig {\begin{figure}[tbhp] \centering}
\def\efig {\end{figure}}

\normalsize \baselineskip=8mm  \vspace{15mm}


\begin{abstract}

In a search for a simple proximity system of a topological insulator and a superconductor for studying the role of surface versus bulk effects by gating, we report here on a first step toward this goal, namely the choice of such a system and its characterization. We chose to work with thin film bilayers of grainy 5 nm thick NbN films as the superconductor, overlayed with 20 nm thick topological layer of $\rm Bi_2Se_3$ and compare the transport results to those obtained on a 5 nm thick reference NbN film on the same wafer. Bilayers with ex-situ and in-situ prepared $\rm NbN-Bi_2Se_3$ interfaces were studied and two kinds of proximity effects were found. At high temperatures just below the superconducting transition, all bilayers showed a conventional proximity effect where the topological $\rm Bi_2Se_3$ suppresses the onset or mid-transition $T_c$ of the superconducting NbN films by about 1 K. At low temperatures, a cross-over of the resistance versus temperature curves of the bilayer and reference NbN film occurs, where the bilayers show enhancement of $T_c(R=0)$, $I_c$ (the supercurrent) and the Andreev conductance, as compared to the bare NbN films. This indicates that superconductivity is induced in the $\rm Bi_2Se_3$ layer at the interface region in between the NbN grains. Thus an inverse proximity effect in the topological material is demonstrated.

\end{abstract}

\maketitle

\section{Introduction }
\normalsize \baselineskip=6mm  \vspace{6mm}

Topological superconductors (TOS) are interesting due to the expected zero energy Majorana states at vortices \cite{KaneRMP}, and their potential application in quantum computing \cite{Kitaev}. Bulk TOS such as copper doped $Bi_2Se_3$ should have been the simplest materials to study TOS properties, but complications due to their inherent inhomogeneity \cite{Kriener} and the presence of possible superconducting impurity phases such as $CuSe_2$ \cite{AndoRev}, make them less attractive for such investigations. Another way for investigating TOS is by inducing superconductivity in a topological insulator or in semiconductor-nanowires with strong spin-orbit coupling via the proximity effect (PE) \cite{Koren1,LiLu,Kouwenhoven,Heibloom}. Unconventional superconductivity in these systems, such as reflected by the observation of zero bias conductance peaks (ZBCP), indicates the presence of zero (or near zero) energy bound states that might be due to Majorana zero energy modes. These Majorana modes are strictly located at zero energy while standard Andreev bound states (ABS which also show ZBCPs) can originate also in very close to zero energy states. Thus reducing the thermal broadening of the ZBCPs by further lowering the temperature should help to distinguish between them, but even then, ABS at strictly zero bias could not be ruled out \cite{Xu}. Another complication involves the spatial sharpness of the boundary region between the superconductor and the topological or semi-conducting material. It was shown that a smoothly varying boundary leads to near-zero-energy end states even in the topological trivial case \cite{Brouwer}. Since this boundary is generally created by gating in the experiments, its smoothness and tapering add more uncertainty to the interpretation of the observed ZBCPs as due to Majorana modes \cite{Kouwenhoven,Heibloom}. The role of gating in these nanowire-superconductor experiments was further investigated recently and a variety of additional phenomena such as ZBCP oscillations versus gate voltage and magnetic field were observed and interpreted in the context of Majorana modes as well as alternatives such as Kondo and disorder effects  \cite{Churchill}. Hence, gating is sufficiently important in these studies and we decided to look for a simple proximity system of a topological insulator and a 2D superconductor for studying the role of surface versus bulk effects by gating. We extended our previous investigations of $Bi_2Se_3-NbN$ junctions \cite{Koren2,Koren3} to bilayers of this system using ultra-thin, grainy $NbN$ layers with weak-links in between the grains, overlayed with a thicker $Bi_2Se_3$ layer which facilitates stronger-links between the $NbN$ grains via the (inverse) proximity effect. Here we report on transport characterization of this system, while gating studies of these bilayers will be performed in the future.\\

\begin{figure} \hspace{-20mm}
\includegraphics[height=7cm,width=8cm]{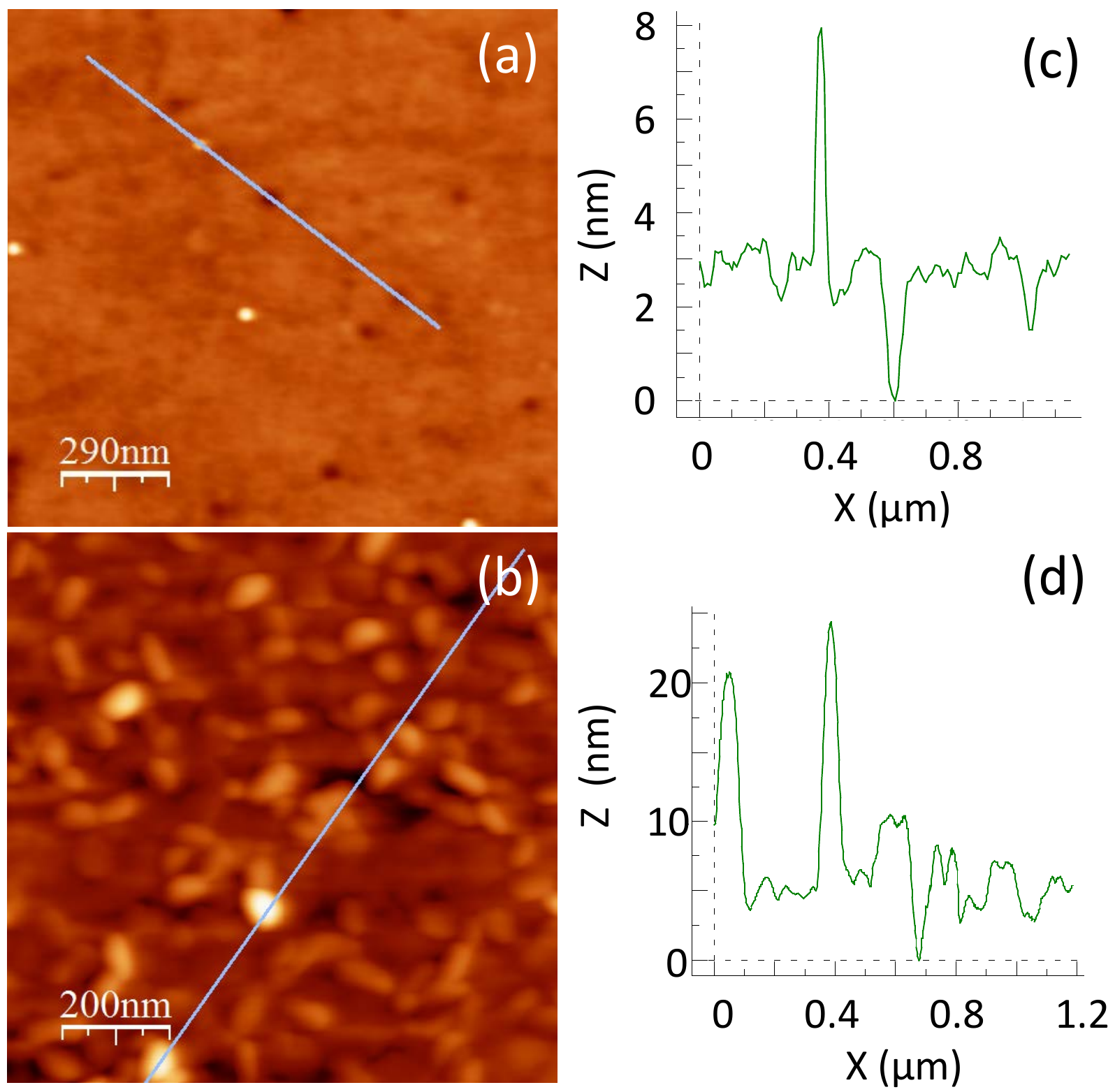}
\hspace{-20mm} \caption{\label{fig:epsart} (Color online) Atomic force microscope (AFM) images of a 5 nm thick NbN film (a) and of a 20 nm  $Bi_2Se_3$ on a 5 nm NbN bilayer (b) together with their corresponding line profiles along the lines shown in (a) and (b), as depicted in (c) and (d), respectively. }
\end{figure}

\section{Preparation and characterization of the films and bilayers }
\normalsize \baselineskip=6mm  \vspace{6mm}

The NbN and $Bi_2Se_3$ thin films were prepared by laser ablation deposition using the third harmonic of a Nd-YAG laser. The NbN films were deposited using a metallic Nb target under 30-40 mTorr of $\rm N_2$ gas flow at 600 $^0$C heater block temperature, while the $Bi_2Se_3$ layers were deposited using a highly rich Se target (pressed at room temperature with Bi:Se atomic ratio of 1:17), under vacuum and at 300 $^0$C. The laser was operated at a pulse rate of 3.33 Hz, with high fluence for the deposition of the NbN films ($\sim 10\, J/cm^2$) and low fluence for the deposition of the $Bi_2Se_3$ layers ($\sim 1\, J/cm^2$). All films were deposited on fused silica wafers of $10\times10\, mm^2$ area. X-ray diffraction measurements of our typical bilayer of 20 nm thick $Bi_2Se_3$ on 5 nm thick $NbN$ on a fused silica wafer, showed that the $Bi_2Se_3$ cap layer grew with preferential c-axis orientation normal to the wafer with c=2.85 nm. Fig. 1 shows atomic force microscopy (AFM) images of the surface morphology of the as deposited NbN film (a) and the bilayer (b), together with their corresponding typical line profiles (c) and (d), respectively. The RMS roughness of the film is $\sim$0.5 nm and that of the bilayer is of about 2 nm. We note that a 10 nm thick $NbN$ layer on (100) $SrTiO_3$ wafer was much smoother with RMS roughness of only 0.1 nm (not shown). Since grainy thin films are essential for the present study (otherwise a superconducting short will mask all our transport data), we chose to use the fused silica substrates on which the films are much rougher as seen in Fig. 1. SEM images of similar NbN films on glass showing their grainy nature can also be seen in \cite{Xin-kang}.   \\

Two types of bilayers were used in the present study. One using an \textit{"ex-situ"} process where the NbN film was exposed to ambient air for 1 hour after which it was measured under He atmosphere, and then immediately put back into the vacuum chamber for the deposition of the $Bi_2Se_3$ cap layer. The other was an \textit{"in-situ"} process where both  $Bi_2Se_3-NbN$ bilayer and reference $NbN$ film were prepared in the same deposition run without breaking the vacuum, on two halves of the wafer by the use of a shadow mask, similar to that described in \cite{KM}. It is well known that NbN films exposed to ambient air develop within several minutes a surface oxide layer of about 1 nm thickness, while oxidation in between the grains takes longer \cite{Darlinski}. It was thus essential to investigate both ex-situ and in-situ bilayers in order to see how the $NbN_xO_y$ oxides affect the results. Transport measurements were done by the use of an array of 36 gold coated spring loaded spherical tips for the 4-probe measurements on nine different locations on the wafer. Fig. 2 depicts a schematic drawing of an in-situ prepared sample with the bilayer and reference NbN film (separated by a scratch) and the 36 contact locations. Also shown is a representative current and voltage wiring scheme which is switched from contact C1 to contact C10 using an electronic switching box. In the present study, contacts C1, C3, C5 (on the scratched area) and C9 were disconnected, but by reversing the sample position in the measuring probe, most of the shown contact locations could be measured (in reversed order C(i) is C(11-i)).\\

\begin{figure} \hspace{-20mm}
\includegraphics[height=6cm,width=8cm]{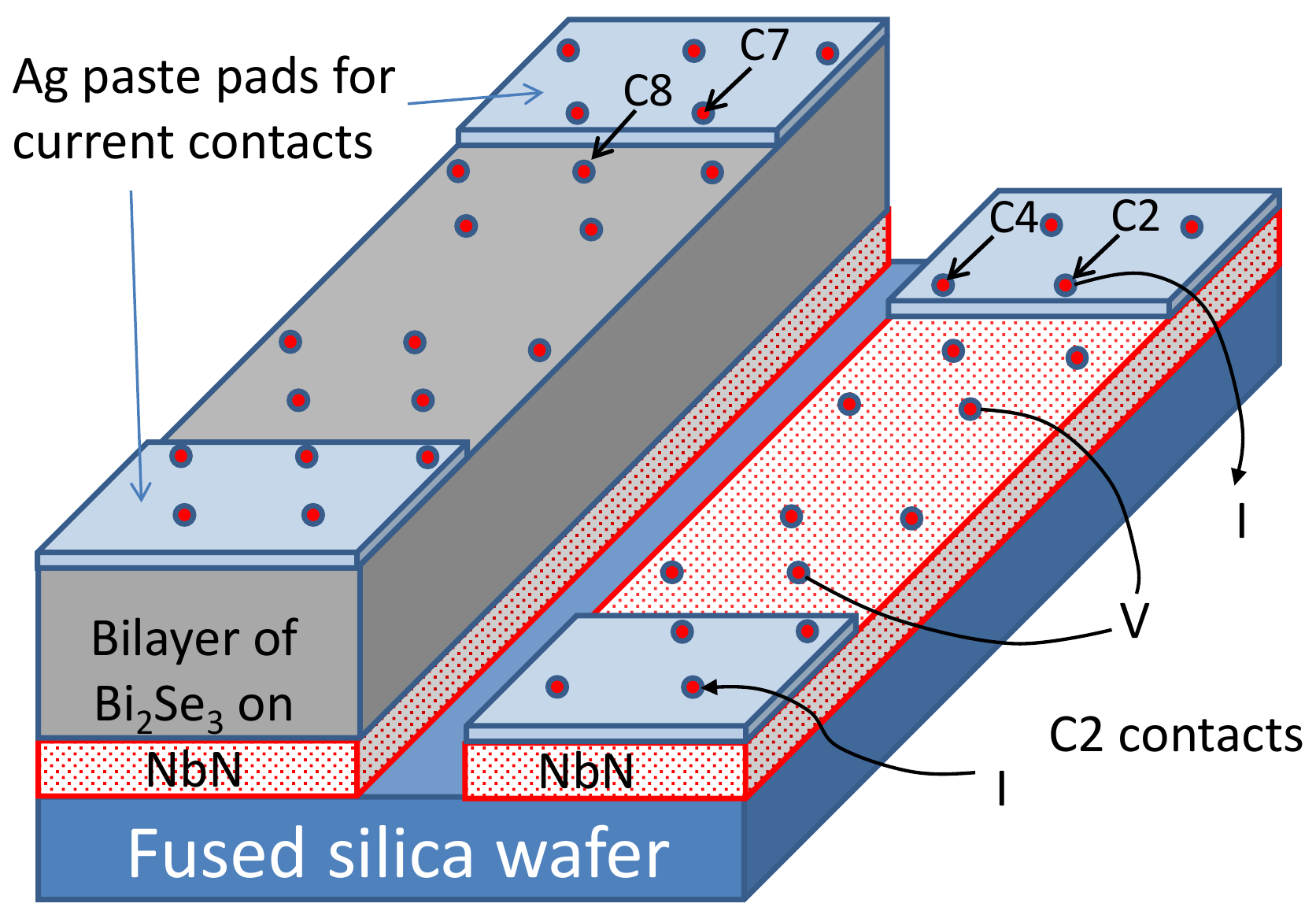}
\hspace{-20mm} \caption{\label{fig:epsart} (Color online) A schematic drawing of an in-situ prepared bilayer of $Bi_2Se_3$ on NbN, and reference NbN film. The geometry of the $9\times4$  current and voltage contacts from C1 to C10 is also shown (C5 that falls on the separation area, is not shown). }
\end{figure}

\section{Results and discussion}

\subsection{\textit{ex-situ} prepared bilayers}

\begin{figure} \hspace{-20mm}
\includegraphics[height=7cm,width=8cm]{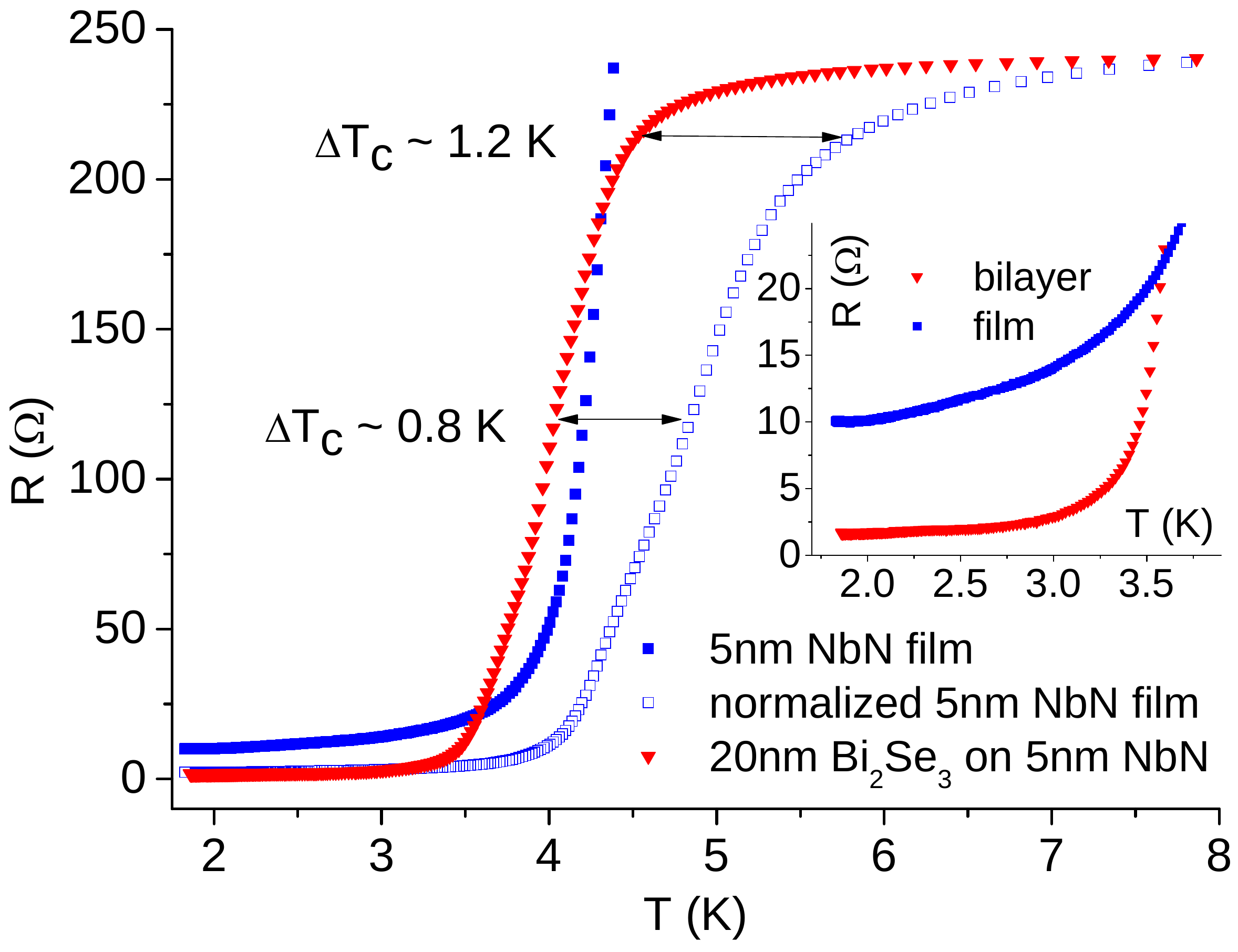}
\hspace{-20mm} \caption{\label{fig:epsart} (Color online) Resistance versus temperature of a 5 nm NbN film after its exposure to ambient air for about 1 hour, and of a bilayer of 20 nm $Bi_2Se_3$ deposited on this film. Also shown is an R versus T curve of the 5 nm NbN film normalized to that of the bilayer at 8 K. The inset shows a zoom-in on low temperatures.  }
\end{figure}

Fig. 3 shows the resistance versus temperature results of a 5 nm thick NbN film that was exposed to ambient air for 1h, together with the results measured on a bilayer created on it by the deposition of additional 20 nm thick $Bi_2Se_3$ cap layer. Also shown is the same R-T data of the NbN film normalized at 8 K to that of the bilayer. One can see that at high temperatures just below the transition, the $Bi_2Se_3$ cap layer suppresses $T_c$ of the NbN film in the bilayer by 0.8 K at mid-transition and by 1.2 K in the onset regime at 0.9R(8 K). This indicates a conventional proximity effect like in normal-superconductor (NS) junctions, where the normal electrons from N penetrate S while suppressing the order parameter in it and therefore also $T_c$. It turns out that the topological insulator here behaves like a normal metal. This is not surprising since there is still a significant bulk contribution to the $Bi_2Se_3$ conductance (in addition to the surface one), though we reduced it by two orders of magnitude as compared to our previous study \cite{Koren1} by the use of a Se rich target. The resistivity of the bilayer in Fig. 3 (5 m$\Omega$cm at 8 K) is mostly due to the 20 nm thick $Bi_2Se_3$ cap layer, which corresponds to an electron density of about $10^{17}$ cm$^{-3}$ \cite{Butch}. At low temperatures, a cross-over of the (un-normalized) resistance curves of the reference film and bilayer occurs at 3.6 K, and at 2 K the ratio of resistances is $\sim$6.2. The corresponding ratio of the normal state resistances at 8 K is 4.6. Thus, the bilayer resistance due to the $Bi_2Se_3$ cap layer in between the NbN grains is reduced by more than expected from the normal resistance values, and this is a sign of a PE in the $Bi_2Se_3$ layer at the interface which will be further investigated in the following.\\

\begin{figure} \hspace{-20mm}
\includegraphics[height=6cm,width=8cm]{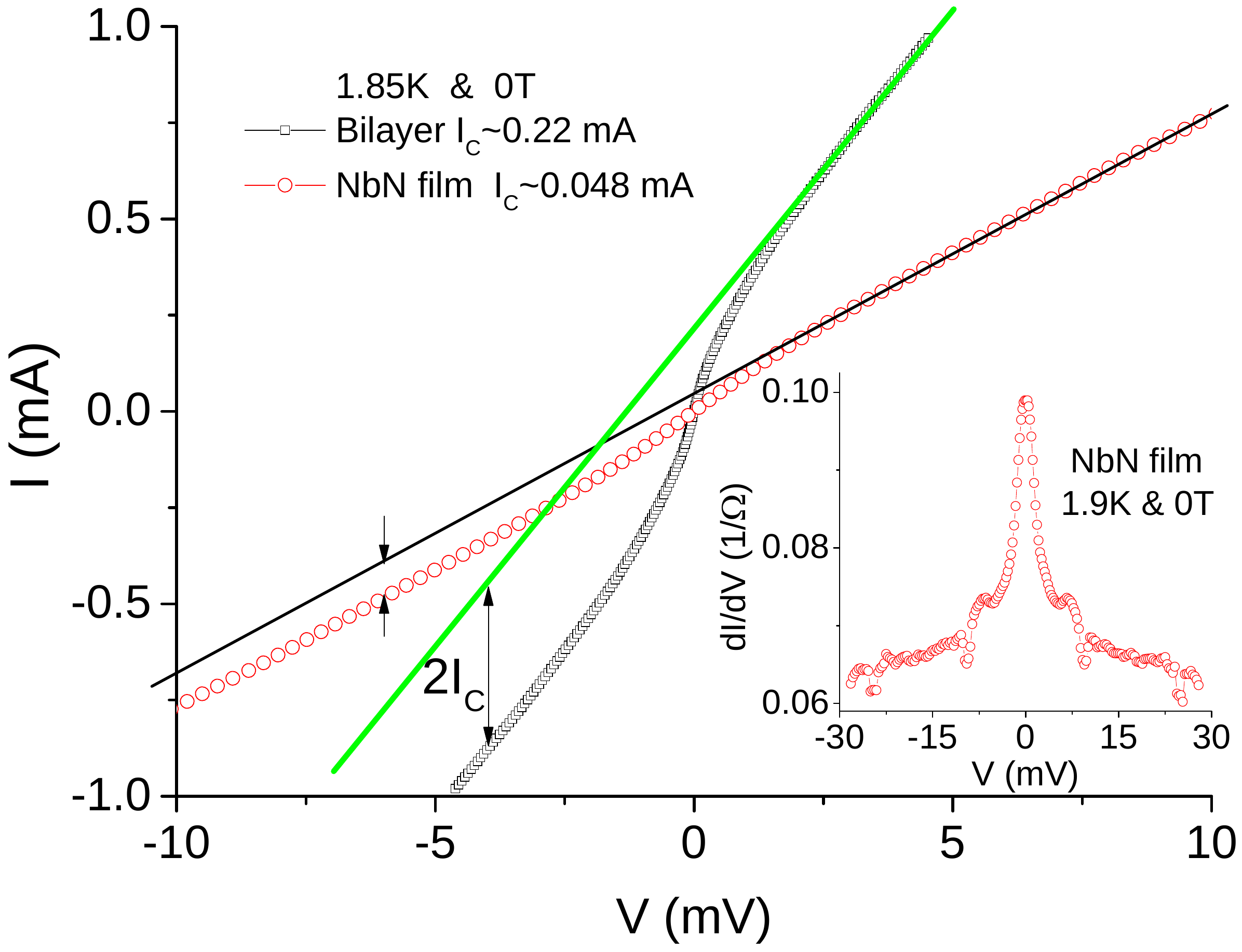}
\hspace{-20mm} \caption{\label{fig:epsart} (Color online) Current versus voltage curves (IVCs) at 1.85 K and zero magnetic field of the NbN film and bilayer of Fig. 3. The inset shows a typical conductance spectrum of the 5 nm thick NbN film.   }
\end{figure}

Fig. 4 shows the I-V curves of the bilayer and NbN film of Fig. 3. Both reveal small critical currents with serial resistance. The inset to Fig. 4 shows a typical conductance spectrum  of the NbN film at 1.9 K and zero field. It reveals a supercurrent peak with finite resistance at low bias, a broader Andreev structure up to about 8 mV, and supercurrent dips \cite{IcDips} at $\pm$10 and $\pm$25 mV. All these demonstrate the weak-link character of the contacts between the NbN grains in this film. The serial resistance could originate in the thin oxide layer in between the grains or the inter-grain resistance in an ultra thin film. The large width of the main peak can be due to supercurrent distribution in the grains, and the broader than the gap Andreev structure to a few weak-links connected in series between the voltage contacts (see Fig. 2). The supercurrent dips are known to originate in heating effects \cite{IcDips}, and here they just show again the presence of supercurrents. Fig. 5 shows the conductance spectra of the bilayer of Fig. 3 at 1.84 K under various magnetic fields normal to the wafer. Here a supercurrent peak which is much narrower now is observed, together with Andreev bound state structures at zero field. With increasing field the main peak decays slowly, while the Andreev structure decays fast and disappears already at a 0.3 T field. Again, this is a signature of weak-links which are stronger in the bilayer than in the film. This effect will be further enhanced in the in-situ bilayers as we shall see next.\\

\begin{figure} \hspace{-20mm}
\includegraphics[height=6cm,width=8cm]{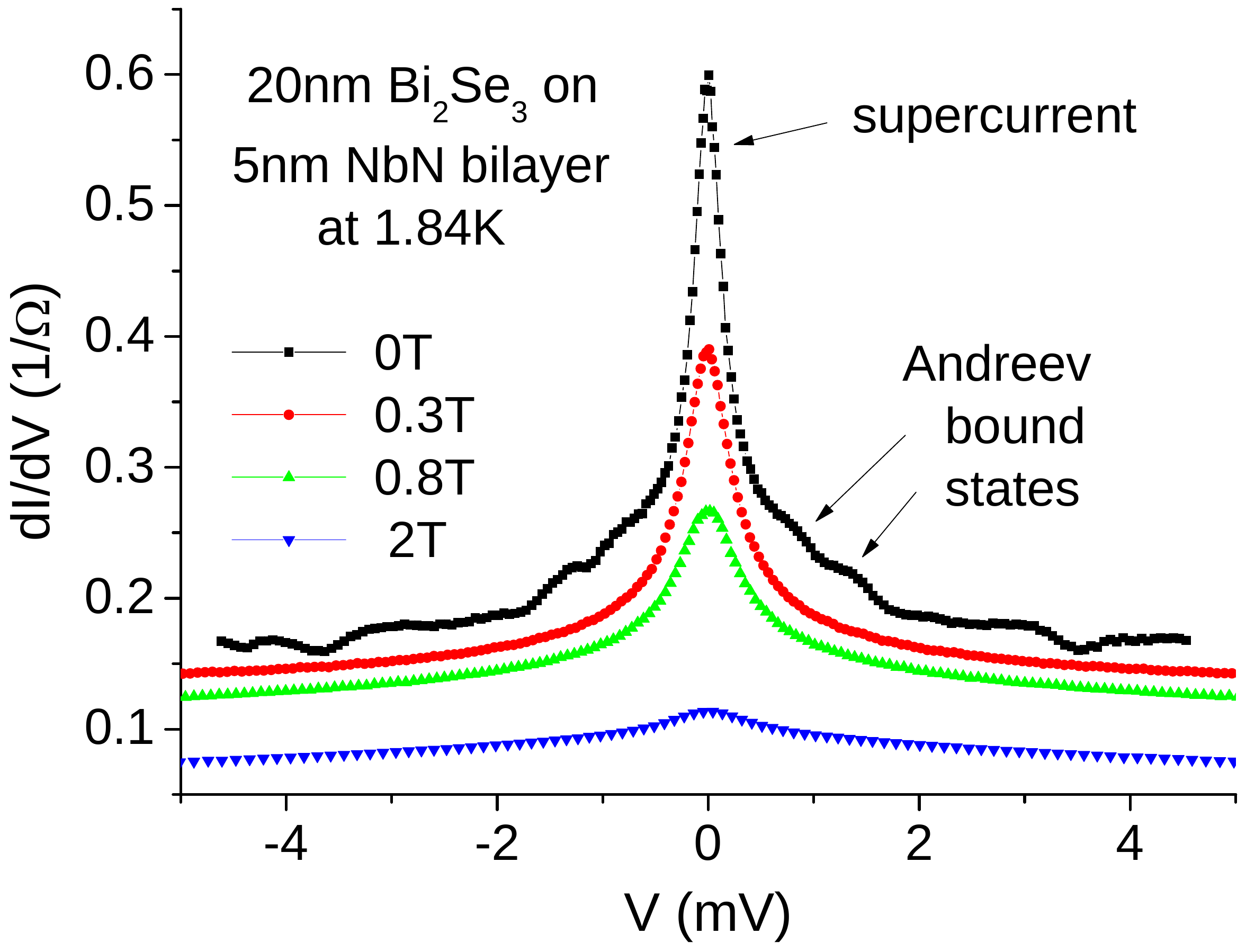}
\hspace{-20mm} \caption{\label{fig:epsart} (Color online) Conductance spectra at 1.84 K at various magnetic fields of the bilayer of Fig. 3. }
\end{figure}

\begin{figure} \hspace{-20mm}
\includegraphics[height=6cm,width=8cm]{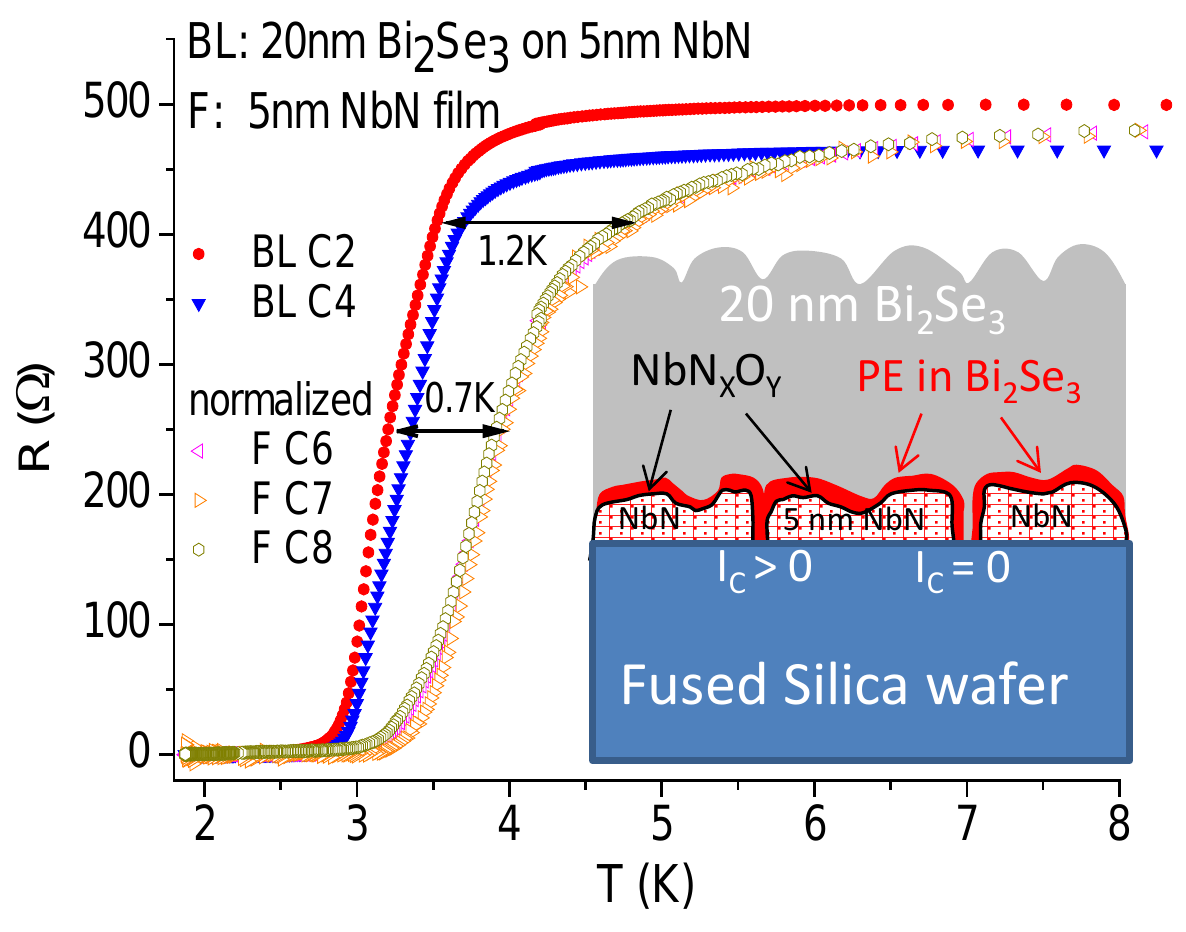}
\hspace{-20mm} \caption{\label{fig:epsart} (Color online) Resistance versus temperature of an in-situ prepared bilayer of 20 nm $Bi_2Se_3$ on 5 nm NbN (shown are the C2 and C4 results) and of the C6, C7 and C8 contacts on the fresh 5 nm NbN reference film normalized to that of the bilayer at 8 K. The inset shows a schematic layout of the bilayer, with the proximity induced superconductive layer in the $Bi_2Se_3$ marked in red. }
\end{figure}

\subsection{\textit{in-situ} prepared bilayers}

Fig. 6 shows the resistance versus temperature of an in-situ prepared bilayer together with that of a fresh reference NbN film normalized to the bilayer resistance at 8 K. To demonstrate the spread of the data on the wafer, we show the results of two locations on the bilayer (on the C2 and C4 contacts, see Fig. 2) and three on the NbN film (C6, C7 and C8). The lower absolute $T_c$ values in Fig. 6 as compared to those of Fig. 3 are due to the higher $N_2$ gas pressure used in the deposition process (40 versus 30 mTorr, respectively). Nevertheless, at high temperatures the same kind of conventional PE is observed where the $T_c$ values of the bilayer are lower by 0.7-1.2 K as compared to those of the reference NbN film. At low temperatures however, the inverse PE in the $Bi_2Se_3$ cap layer is much more pronounced in the in-situ deposited bilayer than in the ex-situ one. This is demonstrated in the main panel of Fig. 7, where the bilayer resistance drops to zero at $T_c\sim$2.3 K while the reference NbN film remains resistive. The noisier data of the NbN film is due to the poor quality of the voltage contacts to this ultrathin and grainy film. The inset to Fig. 7 shows the corresponding I-V curves at 1.82 K of the bilayer and reference film of the main panel. The critical currents of both are much higher here than in Fig. 4, due to the oxide-free interface in the bilayer, and to the minimal oxide layer in the fresh NbN reference film (after about 10 min exposure to ambient air). Fig. 8 shows the conductance spectra of these bilayer (inset) and reference NbN film (main panel). One can see that the supercurrent peaks are much narrower now and the residual resistance is due mostly to flux flow. Moreover, the zero bias conductance of the bilayer is higher by a factor of $\sim$17 as compared to that of the reference NbN film. These observations together with the Andreev behavior at higher bias indicate that strong-links are established in between the grains of the NbN layer in the in-situ bilayers, originating in a strong inverse PE phenomenon. This can be further explained by the inset of Fig. 6, where the scheme shows three separated NbN grains. In the left hand side contact between two of these grains, the proximity regimes in the $Bi_2Se_3$ cap layer overlap thus leading to a strong contact with $I_c>0$. If this kind of contacts percolate between the corresponding voltage contacts, a strong-link behavior is seen. Similar results were obtained previously in the cuprates \cite{OferN,KM2}.\\

\begin{figure} \hspace{-20mm}
\includegraphics[height=6cm,width=8cm]{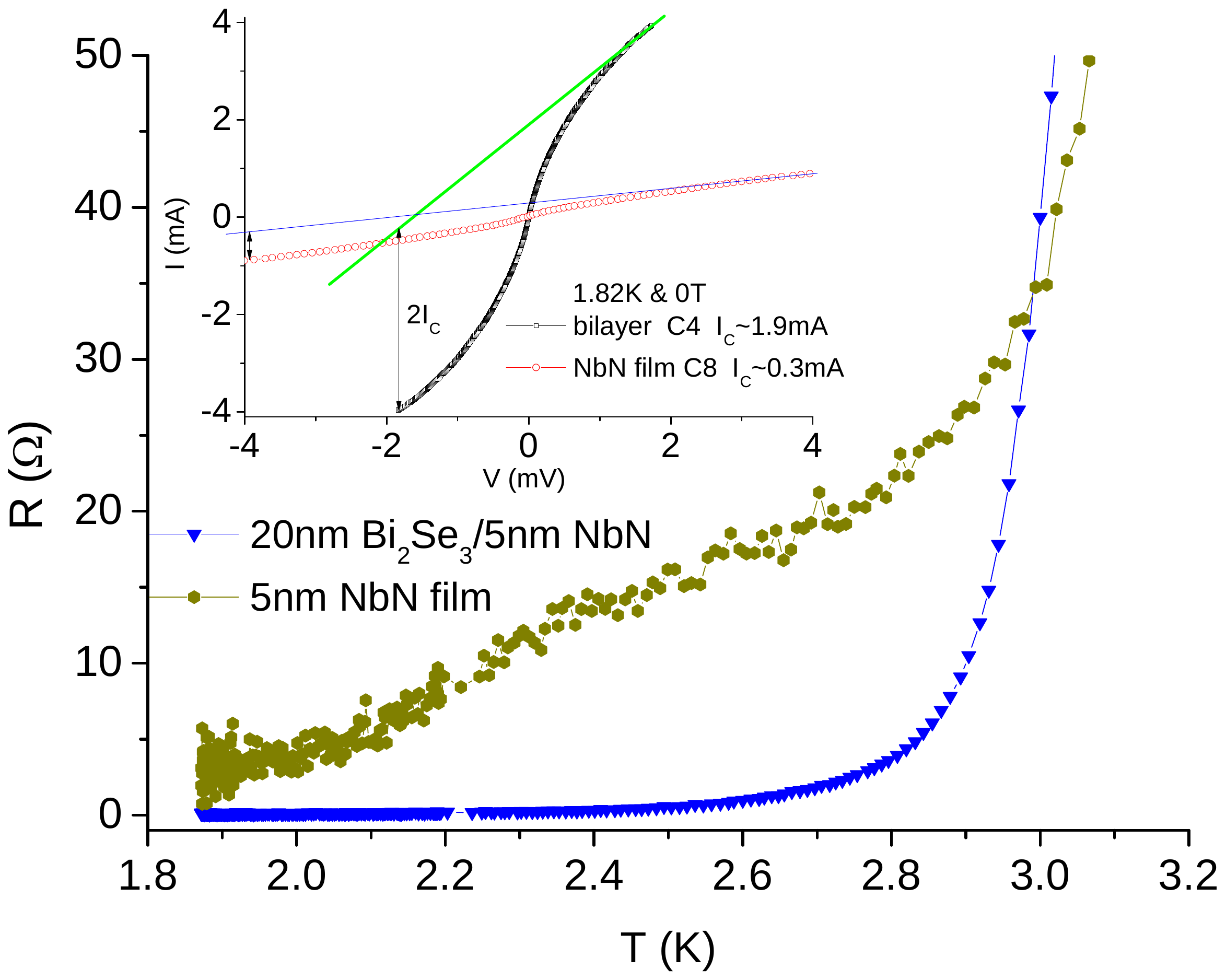}
\hspace{-20mm} \caption{\label{fig:epsart} (Color online) Zoom-in on the low resistance versus temperature regime of the bilayer and reference NbN film of Fig. 6. The inset shows the corresponding IVCs at 1.82 K and zero field.}
\end{figure}

\begin{figure} \hspace{-20mm}
\includegraphics[height=6cm,width=8cm]{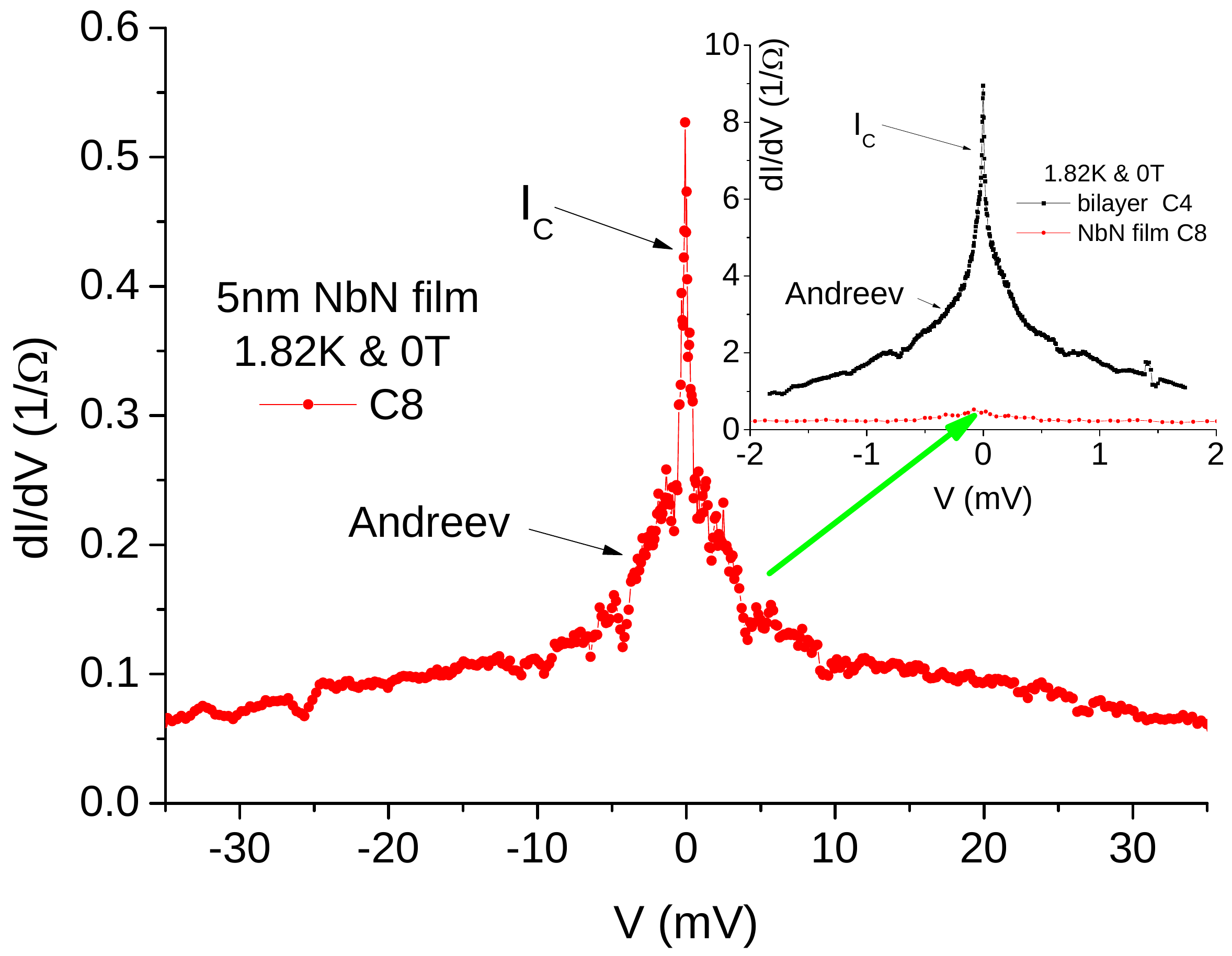}
\hspace{-20mm} \caption{\label{fig:epsart} (Color online) Conductance spectra at 1.82 K  and zero field of the 5 nm NbN film of Fig. 7. The inset shows the conductance of the corresponding bilayer of Fig. 7 together with the data of the main panel for demonstrating its small magnitude.   }
\end{figure}

\begin{figure} \hspace{-20mm}
\includegraphics[height=6cm,width=8cm]{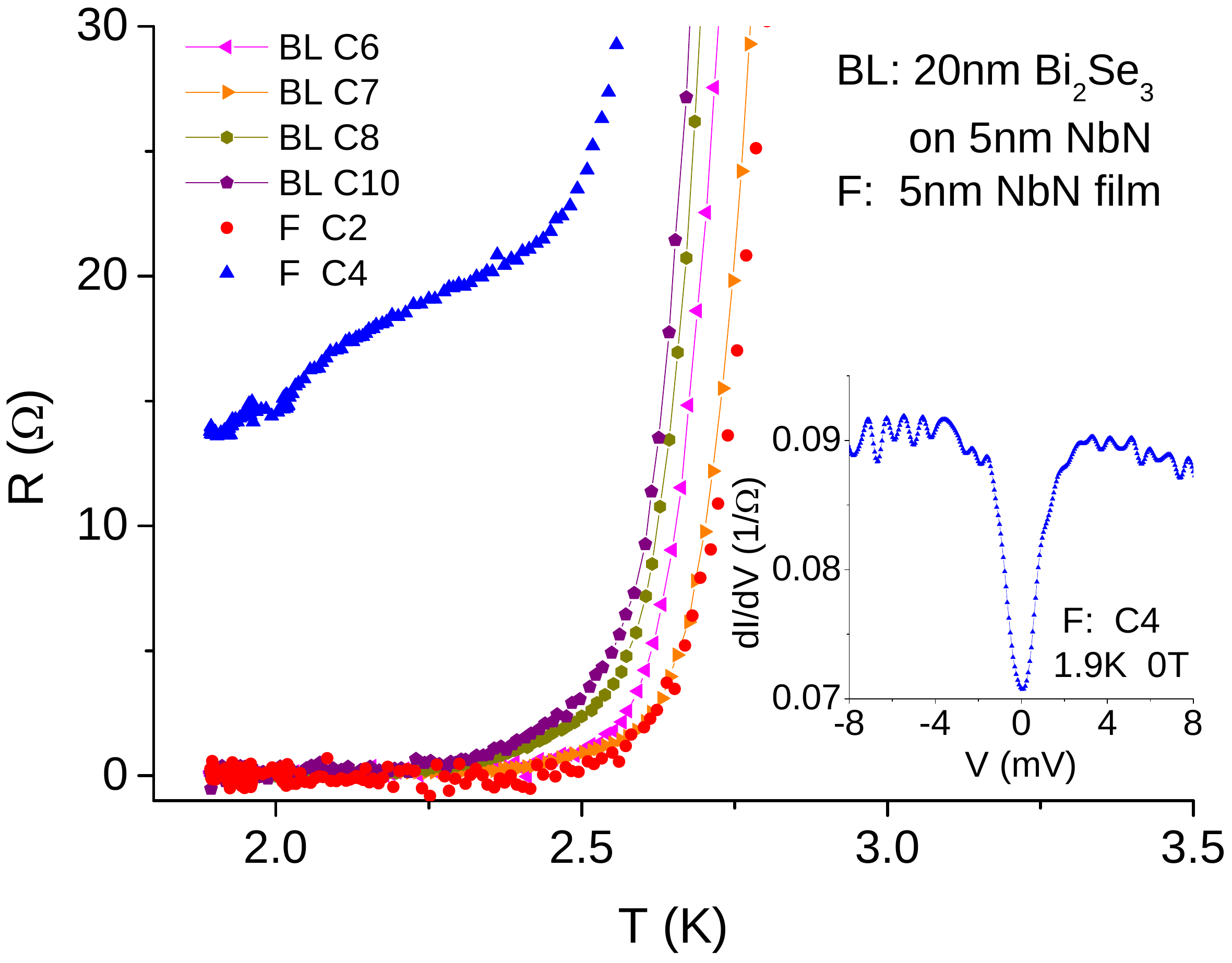}
\hspace{-20mm} \caption{\label{fig:epsart} (Color online) Resistance versus temperature of fresh in-situ prepared $Bi_2Se_3$-NbN bilayer and reference NbN film on a different wafer. The data shows very small scattering of the bilayer results with $T_c(R=0)$ of 2.2-2.3 K, while the reference film in this special and rare case shows a very large scattering of the data with $T_c(R=0)$ of 2.4 K for the C2 contact and no $T_c(R=0)$ at all for the C4 contact. The inset shows the tunneling-like conductance spectrum of the C4 contact.     }
\end{figure}

We shall now focus on scattering of the data on different wafers. As we have seen in Fig. 6, scattering of the data was quite small for both bilayer and reference NbN film. This however, was not always the case as Fig. 9 shows. While the in-situ prepared bilayers with their protected interface have R versus T data very similar to that of Fig. 6, the C2 and C4 contacts on the reference NbN film in Fig. 9 have very different properties. In this quite rare case, C2 is properly superconducting with $T_c$ of about 2.4 K, while C4 is clearly resistive down to 1.8 K. The inset to Fig. 9 shows tunneling-like behavior in the conductance spectrum of the C4 contact, indicating its weak-link behavior at low temperature. The different behaviors of these two contacts reflects the fact that there was a good superconduction percolation path between the NbN grain of the film in C2, but not in C4 where discontinuities (or cuts) between the voltage contacts occurred. These apparently originated in defects in the film in the C4 location, which should have been sub-nano meter in size, as they could not be detected using the atomic force microscope. Referring again to the scheme in the inset to Fig. 6, one can see that the right hand side contact between the NbN grains has no supercurrent ($I_c=0$). Thus, if a series of this kind of contacts cut the superconducting percolation path between the NbN grains, the result is a resistive contact, like C4 in Fig. 9.  \\

\begin{figure} \hspace{-20mm}
\includegraphics[height=6cm,width=8cm]{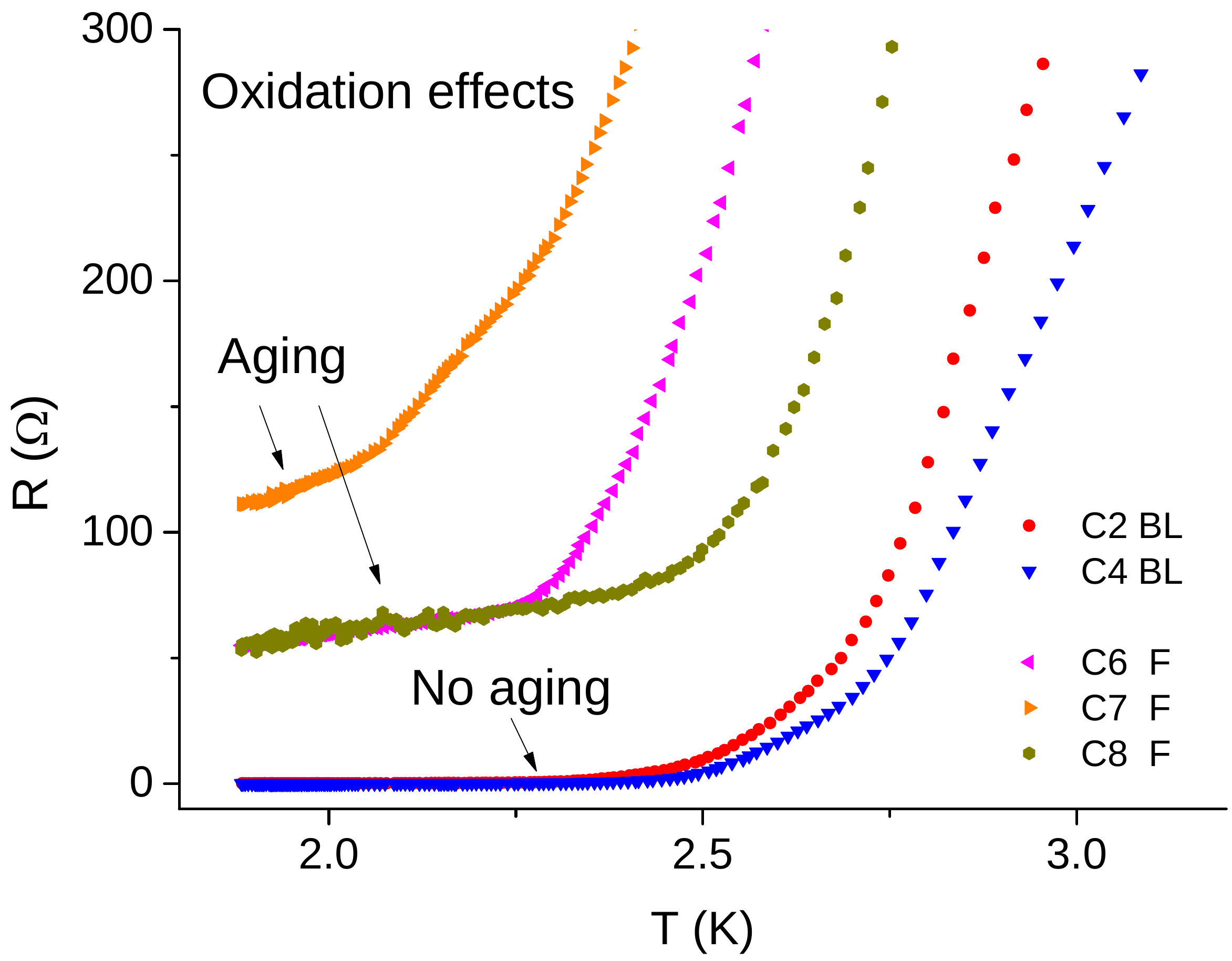}
\hspace{-20mm} \caption{\label{fig:epsart} (Color online) Resistance versus temperature of the in-situ prepared NbN films and bilayer of Fig. 9 after exposure to dry air for 72 hours. Oxidation or "aging" effects are observed here where all the NbN contacts (C6, C7 and C8) are resistive at low T now while the bilayer remains unchanged with $T_c(R=0)$ of 2.2-2.3 K since the interface is protected from oxidation by the ambient dry air. (Measured in reversed order compared to Fig. 9, thus C(i) here is C(11-i) in Fig. 9).   }
\end{figure}

Another parameter that affects our results is oxidation or "aging" in ambient air. Fig. 10 shows this aging effect on the wafer of Fig. 9 after 72 hours exposure to dry ambient air. These R versus T measurements were taken on a reversely inserted wafer, thus the bilayer has the lower contact numbers now, and C7 here is the same contact as C4 in Fig. 9. The R versus T results of the bilayer seem unaffected, with no aging effects of its protected interface layer. All the contacts of the reference NbN film however, become resistive now, with the one that was already resistive in Fig. 9 (C4) having the highest resistance (C7 here). Thus oxidation of the NbN grains in the reference film after a long exposure to ambient air renders the inter grain connections resistive. The bilayer interface though is protected against oxidation and could be used in future gating experiments for longer times.\\

\begin{figure} \hspace{-20mm}
\includegraphics[height=6cm,width=8cm]{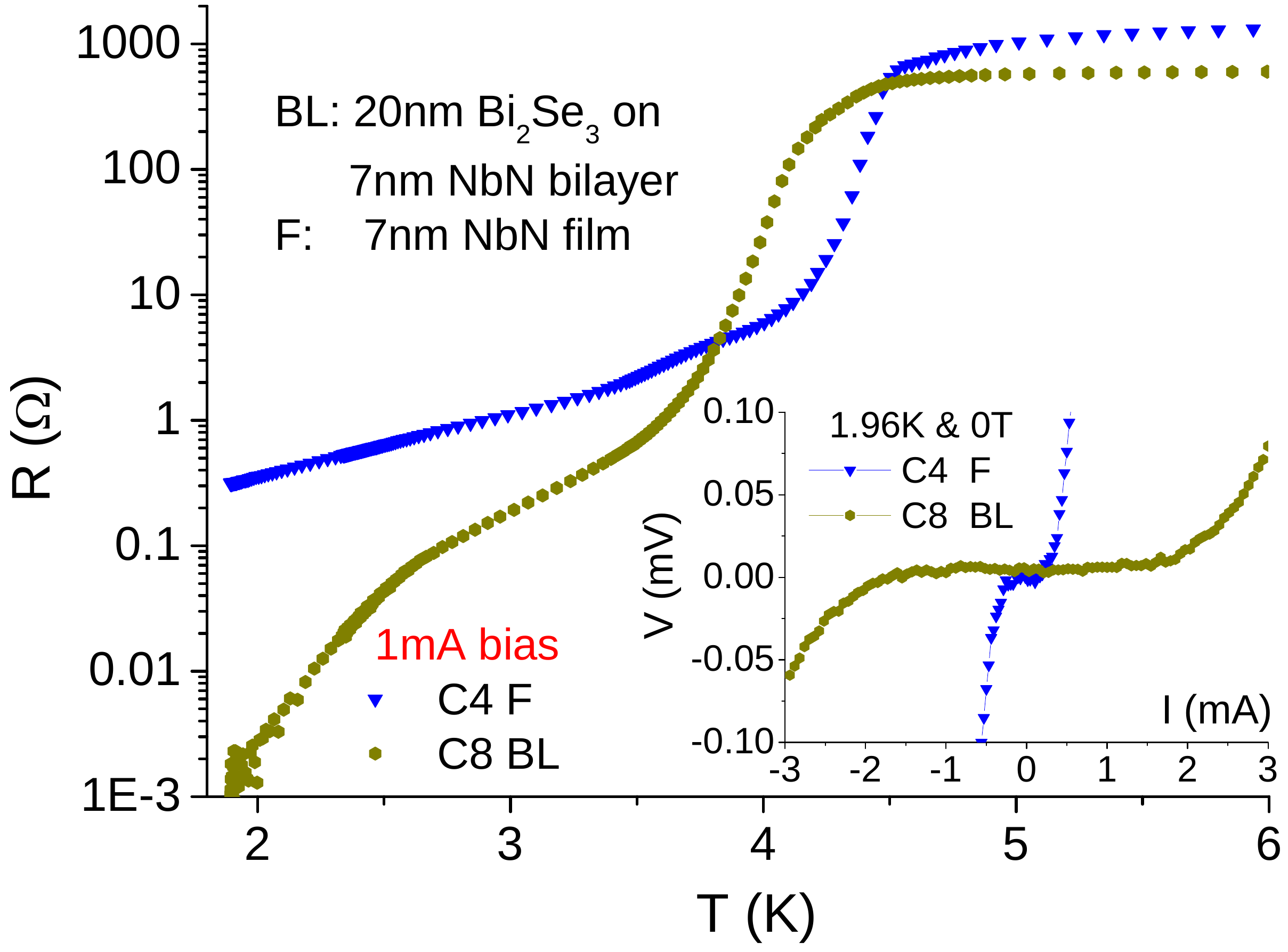}
\hspace{-20mm} \caption{\label{fig:epsart} (Color online) Resistance versus temperature of in-situ prepared 20 nm $Bi_2Se_3$ on 7 nm NbN bilayer and a 7 nm thick NbN reference film using a 1 mA bias current (all previous R-T curves were obtained at low bias of 0.01 mA). The inset shows the corresponding V-I curves. }
\end{figure}

Finally, we describe test results also of bilayers with different NbN layer thicknesses. Using thinner NbN films with 3-4 nm thickness, yielded insulating-like R versus T dependence with only a change of slope at $T_c$ of the grains, similar to what was observed in Ref. \cite{KM2} in the cuprates. So we decided not to work with these films for the PE study. Fig. 11 shows the R versus T results under 1 mA bias current of a bilayer of 20 nm $Bi_2Se_3$ on 7 nm NbN and 7 nm thick reference NbN film. Basically, the same effects as observed before with the 5 nm thick NbN layer were found, but now since the 7 nm thick NbN layer has stronger inter grain links, the residual resistances at low temperatures of both bilayer and reference NbN film are even lower. At high temperatures, the mid transition $T_c$ of the NbN reference film is higher by about 0.3 K than that of the bilayer (conventional PE), while at low T the bilayer resistance is more than two orders of magnitude smaller than that of the reference film. This can be understood by the marked difference between the supercurrent of the two as seen in the inset of this figure. We stress that all the previous R versus T results were obtained using 0.01 mA bias only (compared to 1 mA in Fig 11). In the present case however, under 0.01 mA bias current, both bilayer and reference film had the same $T_c$(R=0) of about 2.5-2.8 K, consistent with the V-I curves in the inset to Fig. 11. Thus, the inverse PE in the $Bi_2Se_3$ cap layer is demonstrated also for these thicker NbN layers. Clearly, if even thicker NbN layers would be used, a superconducting short between the NbN grains would mask any PE effect in the bilayers.\\

\section{Conclusions}

A study of proximity induced superconductivity in bilayers made of topological $Bi_2Se_3$ thin films capping ultrathin superconducting NbN layers was carried out. The transport results of every bilayer were compared to those of its reference NbN film of the same thickness as in the bilayers and prepared on the same wafer. Conventional proximity effect that suppresses superconductivity in the NbN layer of the bilayer was found at high temperatures, just below the superconducting transition. An inverse proximity effect where superconductivity was induced in the topological material was found at low temperatures. We plan to use the present kind of bilayers in future gating experiments. \\

{\em Acknowledgments:}  This research was supported by the Karl Stoll Chair in advanced materials at the Technion.\\

\bibliography{AndDepBib.bib}

\bibliography{apssamp}

\end{document}